\documentclass[twocolumn,prb,showpacs,aps]{revtex4}
\usepackage{amsmath}
\usepackage{amsbsy}
\usepackage{amsthm}
\usepackage{amssymb}
\usepackage{latexsym}

\usepackage[dvips]{color,graphicx}

\begin{document}

\title{Phases and magnetization process of an anisotropic Shastry-Sutherland model }
\author{Zi~Yang Meng and Stefan Wessel}
\affiliation{Institut f\"ur Theoretische Physik III,
Universit\"at Stuttgart, Pfaffenwaldring 57, 70550 Stuttgart, Germany}

\begin{abstract}
We examine
ground state properties of the spin-1/2
easy-axis Heisenberg model on the  Shastry-Sutherland lattice with
ferromagnetic transverse spin exchange using quantum Monte Carlo
and degenerate perturbation theory.
For vanishing transverse exchange, the model reduces to an antiferromagnetic Ising model that besides
N\'eel order harbors regions of extensive ground state degeneracy. In the quantum regime,
we find a dimerized phase of triplet states, separated from the N\'{e}el ordered phase by a superfluid. The quantum phase transitions between these phases are characterized.
The magnetization process shows
a magnetization plateau at $1/3$ of the saturation value, that persists down to the Ising limit, and
a further plateau at $1/2$ only in the quantum regime. For both plateaus, we determine the crystalline patterns of the localized triplet
excitations.
No further plateaus nor supersolid phases are found in this model.
\end{abstract}

\maketitle

\section{Introduction}
Quantum magnets exhibit a wealth of interesting phenomena, in particular on low-dimensional frustrated lattices, where both enhanced quantum fluctuations and geometric frustration  can destroy semi-classical magnetic order. Indeed,
various compounds have been characterized  to provide realizations of the above paradigm.
Recent examples include the valence bond solids found in (C${}_2$H${}_5$)(CH${}_3$)${}_3$P[Pd(dmit)${}_2$]${}_2$~\cite{tamura06a} and
Zn${}_x$Cu${}_{4-x}$(OD)${}_6$Cl${}_2$~\cite{lee07a}.
Another material, that has been intensively studied is SrCuB${}_2$(BO${}_3$)${ }_2$ (we refer to
Ref.~\onlinecite{miyahara03a} for a detailed review of the various experimental and theoretical  explorations on this system).
This compound is described well by the dimer singlet ground state proven exactly previously by Shastry and Sutherland to exist~\cite{shastry81} in the spin-1/2 Heisenberg model on  the orthogonal dimer lattice, shown in Fig. 1.
Recently, new results on the magnetization process of SrCuB${}_2$(BO${}_3$)${ }_2$ have been presented. In particular,  magnetization plateaus at $1/5$, $1/6$, $1/7$,  $1/9$, and $2/9$ of the saturated magnetization have been reported~\cite{sebastian07}, in addition to the previously established plateaus at $1/8$, $1/4$ and $1/3$. While the existence of some of the reported plateaus is at the moment controversial~\cite{levy08}, new B${}^{11}$ NMR data ~\cite{takigawa07} and magnetic torque measurements~\cite{levy08} provide
evidence in favor of a persistent crystalline structure of magnetic excitations also above the $1/8$ plateau. The presence of intra-dimer Dzyaloshinskii-Moriya interactions in  SrCuB${}_2$(BO${}_3$)${ }_2$
however calls for a more complex scenario than
a direct interpretation in terms of supersolidity of triplet excitations in this regime~\cite{levy08}.

As another realization of the Shastry-Sutherland geometry
the rare earth tetraborid TmB${}_4$~\cite{gabani07,siemensmeyer07} has recently been studied in finite magnetic fields.  In contrast to SrCuB${}_2$(BO${}_3$)${ }_2$, this metallic compound exhibits stable long-range antiferromagnetic order in zero field below about 9.8 K. Since full saturation can be obtained for magnetic fields parallel to the c-axis above 5T, TmB${}_4$
allows for a complete scan of its magnetization process. For this compound magnetization plateaus have been observed e.g. at fractions 1/2, 1/7, 1/8 and 1/9 of magnetic saturation. Despite the metallic nature of TmB${}_4$, its magnetism has been suggested to realize an easy-axis anisotropic version of the Shastry-Sutherland model close to the Ising limit with similar intra- and inter-dimer
coupling strengths~\cite{siemensmeyer07}.
\begin{figure}[t]
\centering
\includegraphics[width=0.4\textwidth,height=0.2\textheight]{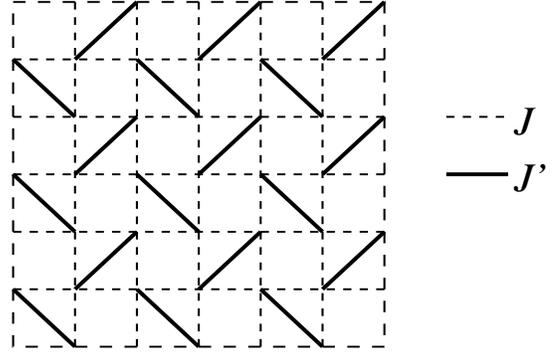}
\caption{The orthogonal dimer lattice of the Shastry-Sutherland model with spin-1/2 degrees of freedom on the square lattice
vertices, and intra-dimer coupling $J'$ (solid lines) and nearest neighbor (inter-dimer) coupling $J$ (dashed lines).}\label{fig:model_1}
\end{figure}

In light of the progress in realizing novel quantum  phases in frustrated quantum magnets, it is important to explore in detail the interplay between geometric frustration and quantum fluctuations in such systems based on effective spin models. In many aspects, numerical studies have become especially important as an unbiased approach to quantum magnetism.
However,
numerical  studies of even simple models of frustrated spin systems  suffer from severe restrictions on the finite sizes accessible to current simulation techniques. In particular, quantum Monte Carlo (QMC) simulations are  tampered by a notorious sign-problem~\cite{troyer05a} due to odd-length spin-exchange paths appearing on non-bipartite lattices. This usually restricts unbiased numerical studies to the small lattices accessible to exact numerical diagonalization. Noteworthy in this respect are however recent studies employing the density matrix renormalization group algorithm on the triangular and kagome lattice Heisenberg model~\cite{white07,jiang08}.

Here, we employ a different approach in order to explore the interplay between quantum fluctuations and frustration,  by studying a model of quantum magnetism in a parameter regime, where geometric frustration is restricted to the classical sector, and does not lead to QMC sign problems.
This allows us to employ large-scale QMC simulations to study quantum effects on a frustrated spin system. In particular,
we study the ground state properties of the spin-1/2 easy-axis Heisenberg model on the orthogonal dimer lattice considered by Shastry and Sutherland~\cite{shastry81}. Namely,  we consider the XXZ Hamiltonian
\begin{eqnarray}
 H=J&\sum_{\langle i,j \rangle}&\left[ -\Delta(S^x_i S^x_j+S^y_i S^y_j)+ S^z_i S^z_j\right]\nonumber\\
  +J'&\sum_{\langle\langle i,j \rangle\rangle}& \left[-\Delta(S^x_i S^x_j+S^y_i S^y_j)+ S^z_i S^z_j\right],
\end{eqnarray}
a variant of the model considered in Ref.~\onlinecite{shastry81} with ferromagnetic transverse exchange ($\Delta>0$) and antiferromagnetic Ising exchange interactions $J,J'\geq0$. Here,
${\mathbf S}_i$ denotes a spin-1/2 degree of freedom on site $i$ of the square lattice, and the first sum extends over all nearest neighbor bonds. The second sum runs over a staggered subset of the next-nearest neighbor bonds, as indicated in Fig. 1.

The model in Eq.~(1) maintains the frustrated nature of the antiferromagnetic Ising interactions, and introduces ferromagnetic spin exchange terms. Employing the well known mapping between
spin-1/2 degrees of freedom and hard-core bosons, the
model can be mapped  onto  an extended bosonic Hubbard model of hard-core bosons  hopping along the bonds of the Shastry-Sutherland lattice and experiencing repulsive interactions proportional to the strength of the Ising exchange.
Recently, similar hard-core boson models have been studied on different lattice geometries, and were found to exhibit interesting order-by-disorder phenomena when quantum fluctuations lift an extensive ground state degeneracy from the Ising limit $\Delta=0$, with new quantum phases emerging. Examples include a supersolid phase on the triangular lattice~\cite{murthy97,wessel05a,melko05,heidarian05},
valence-bond-solids~\cite{isakov06a,damle06a} and a $Z_2$ spin liquid~\cite{balents02a,seng05a,isakov06b} on the kagome lattice, and a $U(1)$ liquid  on  the pyrochlore lattice~\cite{banerjee08}. In the limit of dominating kinetic terms, such models stabilize a superfluid phase on both bipartite and non-bipartite lattices. In magnetic language,  the superfluid corresponds to a transverse ferromagnetic spin alignment, driven by the ferromagnetic nature of the transverse spin exchange.
For the remainder of the paper, we prefer using the spin language, but occasionally find it convenient to also employ the bosonic notation.

As  reviewed in the following section,  the antiferromagnetic Ising  model on the Shastry-Sutherland lattice exhibits regions of extensive ground state degeneracy, similar to the Ising model on the triangular and kagome lattices. Motivated by the above mentioned studies on these frustrated geometries,
we here assess the effects of quantum fluctuations on the classical degenerate ground states on the Shastry-Sutherland lattice, and explore the phase diagram of the full quantum model. We find in this system a dimer triplet state, discussed in detail below, to emerge out of the classical degenerate region. In addition, the system shows a N\'eel  ordered phase and a superfluid regime. We  study the quantum phase transitions  between these different phases, and consider the effects of a magnetic field. We do not obtain indications for supersolidity in this model, but  find that quantum effects lead to the stabilization of a magnetization plateau at 1/2 of the full saturation, that does not persist down to the Ising limit. This is in contrast to the case of e.g. the triangular and kagome lattice, where all plateaus found in the quantum regime persist down to the Ising limit, where they have a largest extension.

The remainder of the paper is organized as follows: In the next section, we review the properties of the antiferromagnetic Ising model on the Shastry-Sutherland lattice. Then, we present in Sec. III our numerical results on the phase diagram of the model introduced above.
In order to explain in a simple picture  the emergence of the dimer triplet phase, we employ degenerate perturbation theory around the Ising limit, which will be discussed in Sec. IV. In Sec. V, we analyze the properties of the system in finite magnetic fields, discuss the appearing  magnetization plateaus, and scan for supersolid phases. Finally, we conclude in Sec. VI by relating our numerical finding to the properties of the isotropic Heisenberg antiferromagnet ($\Delta=-1$) on the Shastry-Sutherland lattice, and discuss connections to recent studies on its magnetization process. We also comment on the recent work on the compound TmB${}_4$, suggested to realize the easy-axis antiferromagnetic Shastry-Sutherland model close to the Ising limit ($-1\ll\Delta<0$)~\cite{siemensmeyer07}.

\section{ Ising Limit}
Before exploring in detail the phase diagram of the quantum spin model introduced above, it is convenient to  review  the properties of the Ising limit, $\Delta=0$, discussed in Ref.~\onlinecite{shastry81}.
\begin{figure}[t]
\centering
\includegraphics[width=0.45\textwidth]{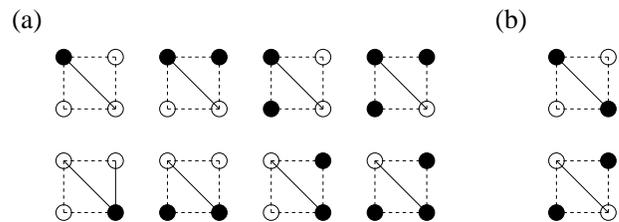}
\caption{(a) Possible spin configurations in the Ising limit on a non-void plaquette for $J'>2J$. (b)  Additional configurations
allowed on a non-void plaquette at $J'=2J$. Full (open) circles denote spin up (down) states.}
\end{figure}
In the Ising limit, the model in Eq.~(1) stabilizes an antiferromagnetic N\'eel phase  for sufficiently weak $J'$, up to  $J'/J<2$. For $J'>2J$ the classical system has a macroscopically degenerate ground state manifold with an extensive ground state entropy $S=[\ln(2)/2] k_B N=0.347  k_B N$, from all configurations that cover each of the $J'$ dimer bonds with
a pair of opposite spins. Here,  $N$ denotes the number of spins. Exactly for $J'=2J$, the degeneracy of the ground state manifold is further enlarged, as additional low-energy configurations proliferate. Shastry and Sutherland proved  a lower bound $S \ge 0.4812 \ k_{B} N$ on the ground state entropy at $J'=2J$, via mapping the model to  a 10-vertex model and using  brading techniques~\cite{shastry81}.
A simple estimate of the ground state degeneracy can be obtained by employing the argument from Pauling's estimate of the residual entropy of ice~\cite{pauling86}. For this purpose, consider one of the filled plaquettes on the Shastry-Sutherland lattice. While for $J'>2J$  the eight configurations shown in  Fig. 2 (a) provide minimal contributions of this plaquette to the total energy, for $J'=2J$  the two configurations shown in  Fig. 2 (b) also lead to a minimal energy contribution.
Given that out of the 16 possible configurations of the four spins forming the plaquette these 10 configurations are thus feasible, we obtain an estimate for the ground state entropy
\begin{equation}
  S \approx k_{B}\ln\bigg[2^{N}\bigg(\frac{10}{16}\bigg)^{N/2}\bigg] = 0.458 \ k_{B} N,
\end{equation}
to be compared to the above bound by Shastry and Sutherland.
We note, that
for $J'>2J$ the Pauling estimate $S = k_{B}\ln(2^{N/2})$
recovers the exact result.

Besides the Ising limit,
Shastry and Sutherland considered the effects of antiferromagnetic transverse spin exchange terms ($\Delta<0$ in our notation), and  proved that the system possesses  an exact dimer-singlet product eigenstate, that  at least for $J'>2J$ becomes the system's ground state~\cite{shastry81}. Later studies by various  groups considered the full quantum phase diagram of this model, which up to date is not conclusively established (c.f. Ref.~\onlinecite{miyahara03a} for a review of the various theoretical and numerical proposals), even though numerical  evidence has been put forward, that the $SU(2)$ symmetric model ($\Delta=-1$) features indeed three phases: (i) a low-$J'$ antiferromagnetically ordered N\'eel phase, (ii) the large-$J'$ dimer singlet phase, and (iii) an intermediate valence bond crystal (VBC) phase, which breaks the lattice symmetry by forming resonating plaquette singlet states on one of the subsets of the void plaquettes of the Shastry-Sutherland lattice~\cite{laeuchli02}. This VBC phase  has a two-fold degenerate ground state and a finite spin excitation gap.
For the remainder of this work, we study the quantum model in the region $\Delta>0$, where large-scale QMC simulations are possible, in contrast to the previously studied case of $\Delta<0$. Furthermore, this model relates directly to a model of hard-core bosons, as mentioned in Sec. I.

\section{Quantum Phase Diagram}
In this section, we present our numerical results on the  phase diagram of the model in Eq. (1). These results are based on QMC simulations of finite systems  with up to $N=36\times 36$ lattice sites, using period boundary conditions. In the simulations, we scaled the inverse temperature as $\beta=1/T=8 L / \Delta J$ in order to access ground state properties. Here, $L$ denotes the linear system size. The QMC simulations were performed employing a generalized directed-loop update~\cite{syljuasen02, alet05} in  the stochastic series expansion (SSE) algorithm~\cite{sandvik99b}. For the results obtained on the larger lattices, in particular in finite magnetic fields, we employed a decoupling of the Hamiltonian in plaquette terms, instead of the more conventional bond decoupling for the SSE formulation.

In Fig. 3, we present the ground state phase diagram resulting from our calculations.
\begin{figure}[t]
\includegraphics[width=0.45\textwidth]{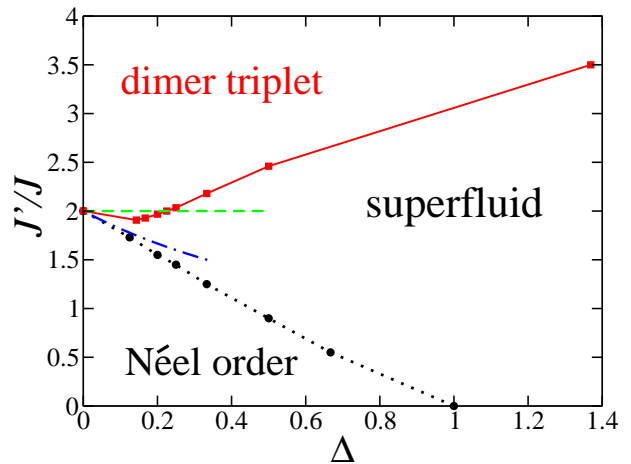}
\caption{(Color online) Ground state phase diagram of the spin-1/2 XXZ model on the Shastry-Sutherland lattice with ferromagnetic transverse spin
exchange. The dotted
(solid) line denotes a first-order (continuous) quantum phase transition. Uncertainties on the indicated phase boundaries are below the symbol size. The dashed line is a guide to the eye indicating a line of constant $J'/J=2$. The dashed-dotted line gives the estimated phase boundary of the dimer triplet phase within perturbation theory around the point $(J'/J, \Delta)=(2,0)$, discussed in Sec. IV.
}
\label{fig:groundstatephasediagram}
\end{figure}
We find that the extend of the antiferromagnetically ordered N\'eel phase shrinks essentially linearly upon increasing $\Delta$ from $0$ up to the Heisenberg point at $(\Delta,J'/J)=(1,0)$ (for $J'=0$, the model reduces to a spin model on the bipartite square lattice, and the sign of $\Delta$ can be inverted by an unitary transformation , thus relating the point $(\Delta,J'/J)=(1,0)$  to the isotropic Heisenberg model at $(\Delta,J'/J)=(-1,0)$).
In hard-core bosonic language, the N\'eel phase corresponds to a checkerboard solid with alternating occupation of the lattice sites. In our QMC simulations, we determine the corresponding structure factor $S_{AF}$ for antiferromagnetic order,
\begin{equation}
 S_{AF}=\frac{1}{N}\sum_{i,j} \epsilon_i \epsilon_j \langle S^z_i S^z_j \rangle
\end{equation}
where $\epsilon_i=\pm1$, depending on the sublattice, to which lattice site $i$ belongs. N\'eel order is present, if in the thermodynamic limit $S_{AF}/N$ scales to a finite value.
For dominant transverse exchange, $\Delta \gg 1$, the model reduces to an ferromagnetic XY model on the Shastry-Sutherland lattice, which in bosonic language relates to an non-frustrated tight-binding hopping model. Hence, the system is expected to exhibit a bosonic superfluid phase for large values of $\Delta>0$, which in spin language relates to a ferromagnetic ordering within the XY plane. Such a phase is characterized by a finite value of the superfluid density, or spin stiffness, which in the QMC simulations can be obtained from measuring the  spin winding number fluctuation~\cite{pollock87} $\langle W^2\rangle$ as
\begin{equation}
\rho_S=\frac{T}{\Delta J} \langle W^2\rangle.
\end{equation}
\begin{figure}[t]
\includegraphics[width=0.45\textwidth]{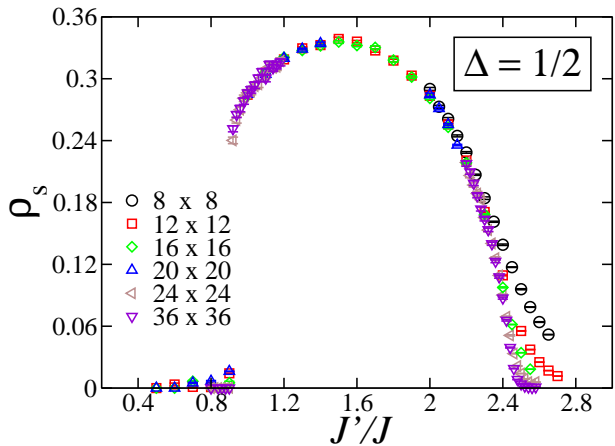}
\caption{(Color online) Spin stiffness $\rho_S$ at fixed $\Delta=1/2$ as a function of $J'/J$ for different system sizes.
}\label{spinstiffness_delta0.5_fullscale.fig}
\end{figure}
As an example, we show in Fig.~\ref{spinstiffness_delta0.5_fullscale.fig} the behavior of $\rho_S$ for different system sizes as a function of $J'/J$ for fixed $\Delta=1/2$. A region with finite spin stiffness is found for $0.9\lesssim J'/J\lesssim 2.5$.
The strong discontinuity of $\rho_S$ at  $J'/J=0.9$ indicates that
the quantum phase transition between the N\'eel ordered phase and the superfluid is strongly first order. Such behavior is also seen by monitoring the antiferromagnetic structure factor $S_{AF}$ upon crossing the phase boundary, as shown in Fig.~\ref{SAF_delta0.5.fig}, again for  $\Delta=1/2$.
Combining the results for $S_{AF}$ and $\rho_S$, we obtain no indication for an intermediate region  exhibiting both finite superfluidity and diagonal long-range order  as inside a supersolid phase, as expected from the commensurate half-filling of the lattice.
\begin{figure}[t]
\includegraphics[width=0.45\textwidth]{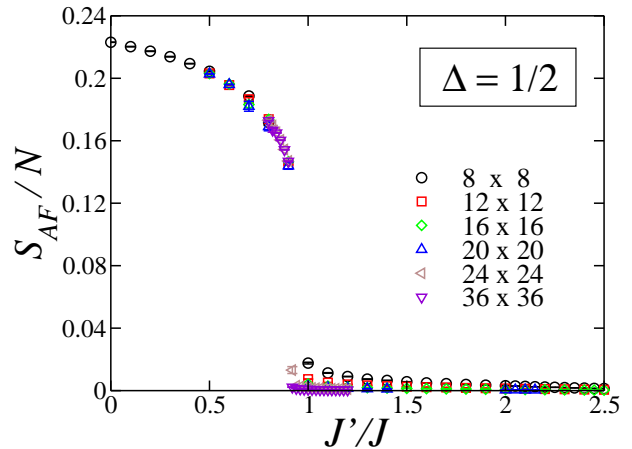}
\caption{(Color online) Antiferromagnetic structure factor across the phase boundary between the N\'eel ordered phase and the superfluid regime at
fixed $\Delta=1/2$ as a function of $J'/J$ for different system sizes.
}\label{SAF_delta0.5.fig}
\end{figure}
For dominant $J'$ (e.g., $J'/J>2.5$ at $\Delta=1/2$), both $S_{AF}$ and $\rho_S$ eventually vanish. We explicitly verified that inside this regime the model does not exhibit  long-ranged correlations in  the  longitudinal nor  the transverse spin-spin correlation function. In addition, also the bond-order-wave structure factors do not exhibit long-ranged order in the spin exchange correlation function (corresponding to kinetic energy correlations in the bosonic model).

\begin{figure}[t]
\includegraphics[width=0.45\textwidth]{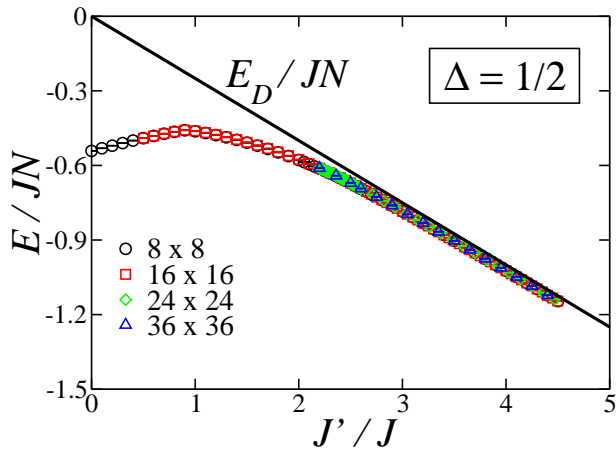}
\caption{(Color online) Comparison of the ground state energy $E$ as a function of $J'/J$ for $\Delta=1/2$ and the variational energy of the
dimerized state of localized $S^z_{tot}=0$ triplet states, $E_D=\langle\psi_D |H|\psi_D\rangle$.
}\label{fig:energycomparison}
\end{figure}

Since the parameter region with $J'>2J$ approaches the degenerate region of the Ising limit for  $\Delta\rightarrow 0$,  quantum effects  indeed select a unique phase from this degenerate ground state manifold. In particular, for small values of $\Delta\ll1$,
the ground state in this large-$J'$ regime can be obtained using degenerate perturbation theory around the Ising limit, discussed in detail in the next section.
Within first-order perturbation theory in $\Delta$ one then finds that for $J'>2J$  quantum fluctuations select the following dimerized state of localized $S^z_{tot}=0$ triplet states on each dimer:
\begin{equation}\label{idealdimernematicstate}
|\psi_D\rangle=\bigotimes_{d} \frac{1}{\sqrt{2}} \left( |\uparrow\downarrow\rangle + |\downarrow\uparrow\rangle\right)_d.
\end{equation}
Here, the direct product extends over all $J'$ dimer bonds on the lattice.
Obviously, this symmetric linear local combination results from the ferromagnetic nature of the transverse spin exchange $(\Delta>0)$ considered here. For $\Delta<0$, one instead recovered the exact dimer singlet state found by Shastry and Sutherland~\cite{shastry81}. In contrast to the dimer singlet state however, the above state  $|\psi_D\rangle$ is not an eigenstate of the Hamiltonian for finite values of $J$.
As discussed in the following sections,  processes in higher order perturbation theory  lead to local correlations between the dominant resonances on the dimers.
Hence, the ground state in the large-$J'$ region of the quantum phase diagram does not take the above direct product form, but approaches it for $J/J'\rightarrow0$. From  the ground state energy, we still find that the
state $|\psi_D\rangle$ provides a appropriate variational state for the true ground state in the large-$J'$ regime. This can be seen even at $\Delta=1/2$, i.e. significantly away from the Ising limit, from a comparison between the system's ground state energy $E$ and the variational energy of $|\psi_D\rangle$,
\begin{equation}
 E_D=\langle\psi_D |H|\psi_D\rangle=-\frac{NJ'}{8}\left(1+2\Delta\right),
\end{equation}
shown in Fig.~\ref{fig:energycomparison}.
Due to  the  dominant  formation of  $S^z_{tot}=0$ triplet states on the dimer bonds, we denote this magnetically disordered phase as a dimer triplet phase.

\begin{figure}[t]
\includegraphics[width=0.475\textwidth]{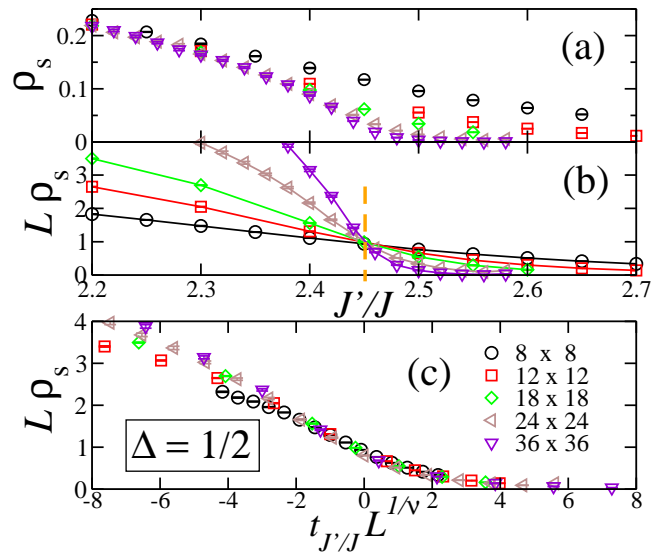}
\caption{(Color online) Spin stiffness $\rho_S$ at fixed $\Delta=1/2$ as a function of $J'/J$ for different system sizes (a), and rescaled with
linear system size $L$ (b), with $J' /J=2.46$ marked by the dashed line.
Part (c) shows the data collapse  expected from a finite size scaling analysis for the 3D $O(2)$ universality class.
}\label{spinstiffness_delta0.5.fig}
\end{figure}

Since no spatial symmetry is broken in the dimer triplet phase, we  expect  the quantum phase transition from the superfluid with broken $U(1)$ symmetry  to the dimer triplet phase to be continuous, and to belong in the universality class of the three-dimensional (3D) $O(2)$ model, with a dynamical critical exponent $z=1$. In order to study the nature of this quantum phase transition in the QMC simulations, we  scanned  the transition region at fixed values  of either $\Delta$ or $J'/J$, varying the other parameter through the phase boundary. Denoting the varied parameter by $X$, at a continuous quantum phase transition the spin stiffness scales
as
\begin{equation}
 \rho_S(X,L)=L^{-z} f \left( t_X L^{1/\nu}, \beta/L^z\right)
\end{equation}
with a scaling function $f$, and the correlations length exponent $\nu$. Here,
\begin{equation}
 t_X=\frac{X-X_c}{X_c}
\end{equation}
denotes the relative distance away from the critical point at $X=X_c$. From the above scaling relation, it follows that $X_c$ can be determined as the crossing point of  finite size data for the rescaled spin stiffness $L^z\rho_S$.
Furthermore, with appropriate values of the critical exponents $z$ and $\nu$, the scaling function $f(\cdot, A)$ is then obtained by plotting $L^z\rho_S $ vs. $ t_X L^{1/\nu}$ for a fixed value of $\beta/L^z=A$.
As an example, we consider a scan in $X=J'/J$ at a fixed value of $\Delta=1/2$, for which the finite size data of $\rho_S$ is shown in Fig~\ref{spinstiffness_delta0.5.fig} (a). For $z=1$, we obtain $X_c=2.46(2)$ from a clear crossing point in Fig~\ref{spinstiffness_delta0.5.fig} (b), and a clear data collapse within a finite critical region, taking $\nu=0.6723$ for the 3D $O(2)$ model~\cite{hasenbusch99}, as shown in Fig.~\ref{spinstiffness_delta0.5.fig} (c).

\begin{figure}[t]
\includegraphics[width=0.49\textwidth]{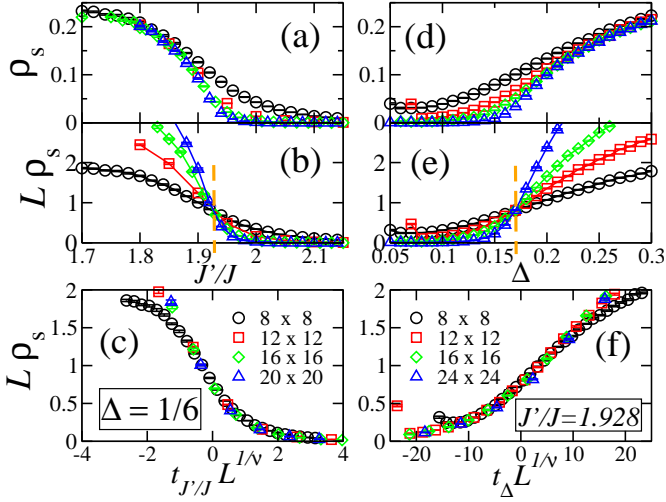}
\caption{(Color online) Left panel: Spin stiffness $\rho_S$ at fixed $\Delta=1/6$ as a function of $J'/J$ for different system sizes (a), and
rescaled with linear system size $L$ (b), with $J'/J=1.928$ marked by the dashed line.
Part (c) shows the data collapse for the 3D $O$(2) universality class. Right panel: Spin stiffness $\rho_S$ at fixed $J'/J=1.928$ as a function of $\Delta$ for different system sizes (d), and rescaled with linear system size $L$ (e), with $\Delta=1/6$ marked by the dashed line.
Part (f) shows the data collapse for the 3D $O$(2) universality class.
}\label{spinstiffness_delta0.167_f1.928.fig}
\end{figure}

\begin{figure}[t]
\includegraphics[width=0.475\textwidth]{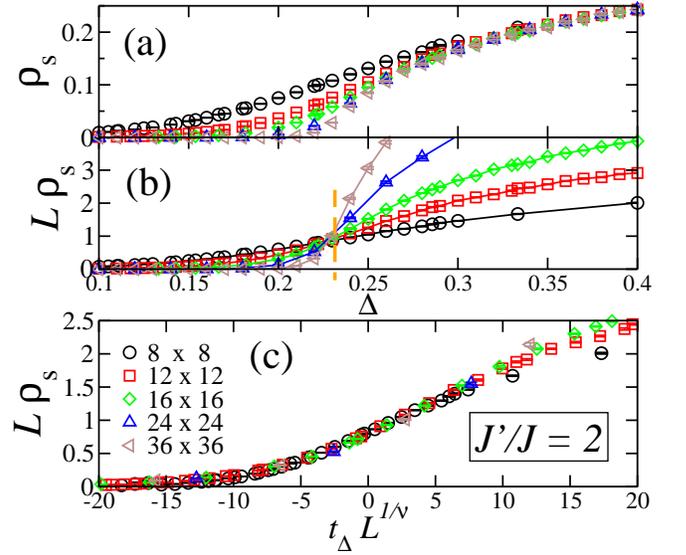}
\caption{(Color online) Spin stiffness $\rho_S$ at fixed $J'/J=2$ as a function of $\Delta$ for different system sizes (a), and rescaled with linear
system size $L$ (b), with $\Delta=0.225$ marked by the dashed line.
Part (c) shows the data collapse  expected from a finite size scaling analysis for the 3D $O$(2) universality class.
}\label{spinstiffness_f2.0.fig}
\end{figure}

Proceeding this way for other values of $\Delta$, we eventually obtained the phase boundary shown in Fig. 3.
From this analysis, we find that at low values of $\Delta\lesssim0.225$, the dimer triplet phase extends below the line $J'/J=2$, indicated by the dashed line in Fig. 3.
In order to illustrate this explicitly, we show in the left panels of Fig.~\ref{spinstiffness_delta0.167_f1.928.fig} the results of the finite size scaling analysis at  $\Delta=1/6$, where the transition point is located at $J'/J=1.928(5)<2$.
As a crosscheck, the right panel of  Fig.~\ref{spinstiffness_delta0.167_f1.928.fig}  shows  the data of the
finite size scaling analysis at fixed $J'/J=1.928$, where the transition is indeed observed at $\Delta=1/6$.
Similarly, when varying $\Delta$ at fixed $J'/J=2$ a transition point between the dimer triplet phase and the superfluid regime is found at $\Delta=0.225(1)$, as extracted from Fig.~\ref{spinstiffness_f2.0.fig}. These results suggest that either (i) the superfluid region separating the N\'eel ordered phase and the dimer triplet phase persists down to the Ising limit, or (ii) the first order transition line and the second order transition line meet at a finite value of $\Delta$, or (iii) an additional phase appears near $J'=2J$ for even smaller values of $\Delta<0.1$.

Such an additional phase might be expected to be selected by quantum effects from the state of
enlarged degeneracy of the Ising model at $J'=2J$ for finite $\Delta$.
The QMC simulations could not be extended to significantly smaller values of $\Delta$, due to an reduced efficiency in parameter regions dominated by the frustrated diagonal part of the Hamiltonian.
However, as  discussed in the following section, degenerate perturbation theory in $\Delta$ indicates that at $J'=2J$ the dominant
effect of a finite $\Delta$ is to effectively drive the system away from $J'=2J$ towards the region $J'>2J$.
This leads us to exclude option (iii) from the above list. Scenario (ii) would imply a direct first order transition between the N\'eel ordered phase and the dimer triplet state for sufficiently low values of $\Delta>0$, whereas within scenario (i) the superfluid phase would always separate the two phases.

\section{Perturbation Theory}
In order to study more closely the emergence of the dimer triplet phase from the Ising limit upon introducing transverse exchange interactions, we employed degenerate perturbation theory, starting from the degenerate ground state manifold in the Ising limit $\Delta=0$. First, we consider the region $J'>2J$, where the degenerate ground state manifold is spanned by independently placing two opposite spins on each $J'$-dimer, as discussed in Sec. II.

\begin{figure}[t]
\centering
\includegraphics[width=0.45\textwidth]{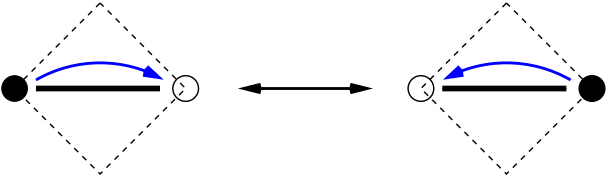}
\caption{(Color online) Spin exchange processes contributing to the effective Hamiltonian $H^{(1)}_{eff}$ in first order perturbation theory. Full (open)
circles denote spin up (down) states.}\label{fig:firstorderperturbation}
\end{figure}

For each such $J'$-dimer $d$, we denote these two lowest energy states as
\begin{eqnarray}
|\Downarrow\rangle_d &=& |\uparrow\downarrow\rangle_d, \nonumber\\
|\Uparrow\rangle_d &=&   |\downarrow\uparrow\rangle_d,
\end{eqnarray}
which
form an effective spin-1/2 degree of freedom on the dimer $d$.
We separate the full Hilbert space of the system  into the model space $M$, spanned by these ground state configurations, and the orthogonal space $O$. A basis of $M$ is given by the orthonormal states
\begin{equation}
|\psi_a\rangle=\bigotimes_d |\psi_{a}\rangle_d,\:\mathrm{with}\:|\psi_{a}\rangle_d\in\{|\Downarrow\rangle_d, |\Uparrow\rangle_d\}.
\end{equation}
The orthogonal space $O$ is spanned by all states of the Ising model, that do not belong to this set.
We denote these orthonormal basis states by $|\psi_b\rangle$, for which at least one dimer $d$ has both spins equal, i.e.
$|\uparrow\uparrow\rangle_d$ or $|\downarrow\downarrow\rangle_d$.
The Hamiltonian of Eq.~(1) is similarly  split into the model Hamiltonian $H_{0}$ and a perturbation part $H_{1}\propto \Delta$,
\begin{eqnarray}
	H_{0} &=& J\sum_{\langle i,j \rangle} S^z_i S^z_j + J'\sum_{\langle\langle i,j \rangle\rangle} S^z_i S^z_j; \nonumber\\
	H_{1} &=& J\sum_{\langle i,j \rangle} \left[ -\Delta(S^x_i S^x_j+S^y_i S^y_j)\right] \nonumber\\
	      & & + J'\sum_{\langle\langle i,j \rangle\rangle}  \left[-\Delta(S^x_i S^x_j+S^y_i S^y_j)\right],
\end{eqnarray}
where $H_0$ is diagonal in the basis of $M$ and $O$ introduced above.
The effective Hamiltonian $H_{eff}$, that describes the effective dynamics induced by $H_{1}$ within the model space $M$ is given by degenerate perturbation theory up to third order in $\Delta$ as~\cite{lindgren74}
\begin{eqnarray}
	H_{eff} &=& PH_{0}P + \underbrace{PH_{1}P}_{\textrm{1st order}} + \underbrace{PH_{1}RH_{1}P}_{\textrm{2nd order}} \nonumber\\
		& & + \underbrace{PH_{1}RH_{1}RH_{1}P - PH_{1}RRH_{1}PH_{1}P}_{\textrm{3rd order}} \nonumber\\
		& & + O(\Delta^4),
\end{eqnarray}
where
\begin{equation}
P = \sum_{a }|\psi_a\rangle\langle\psi_a|
\end{equation}
is the projection operator
onto the model space in terms of the above constructed basis states $|\psi_a\rangle$, and
\begin{equation}
R = \sum_{b}\frac{|\psi_b\rangle\langle\psi_b|}{E_{0} - E^b_{0}}
\end{equation}
is the resolvent operator
with $E_0=-NJ'/8$ the degenerate ground state energy of $H_0$, and $E^b_{0}=\langle\psi_b|H_0|\psi_b\rangle$ the energy of the basis state $|\psi_b\rangle$ of the orthogonal space $O$.
We start by considering the first order contribution to $H_{eff}$. The only possible process in this case
is a single spin-flip along a diagonal bond, graphically represented in Fig.~\ref{fig:firstorderperturbation}.
In terms of effective spin operators $\widetilde{S}^+_{d}, \widetilde{S}^-_{d}$ and $\widetilde{S}^z_{d}$, that act on  the effective dimer spin states $|\Uparrow\rangle_d, |\Downarrow\rangle_d$, the effective Hamiltonian in first order perturbation thus reads,
\begin{equation}
	H^{(1)}_{eff} = -\frac{\Delta J'}{2}\sum_{d}(\widetilde{S}^+_{d} + \widetilde{S}^-_{d}) = -\Delta J'\sum_{d}\widetilde{S}^x_{d},
\end{equation}
corresponding to a uniform transverse magnetic field acting on the effective dimer spins.
The  lowest-energy eigenstate of $H_{eff}$ for $\Delta>0$ is the direct product state
\[
|\psi_D\rangle
=\bigotimes_{d} \frac{1}{\sqrt{2}} \left(|\Uparrow\rangle_d + |\Downarrow\rangle_d\right)
=\bigotimes_{d} \frac{1}{\sqrt{2}} \left( |\uparrow\downarrow\rangle + |\downarrow\uparrow\rangle\right)_d,
\]
referred to already in Eq.~(\ref{idealdimernematicstate}) of Sec. III,
corresponding to the decoupled dimer state with each dimer forming a $S^z_{tot}=0$ triplet state.
In case of an antiferromagnetic transverse exchange, $\Delta<0$, the lowest energy state of $H_{eff}$ is the dimer singlet state
\[
|\psi_S\rangle
=\bigotimes_{d} \frac{1}{\sqrt{2}} \left(|\Uparrow\rangle_d - |\Downarrow\rangle_d\right)
=\bigotimes_{d} \frac{1}{\sqrt{2}} \left( |\uparrow\downarrow\rangle - |\downarrow\uparrow\rangle\right)_d,
\]
proven to be an exact eigenstate of the full Hamiltonian $H$ for $\Delta<0$ by Shastry and Sutherland~\cite{shastry81}.
However, $|\psi_D\rangle$ is not an eigenstate of the full Hamiltonian, and correlations between the effective  dimer spins are introduced in higher order perturbation theory.

\begin{figure}[t]
\centering
\includegraphics[width = 0.45\textwidth]{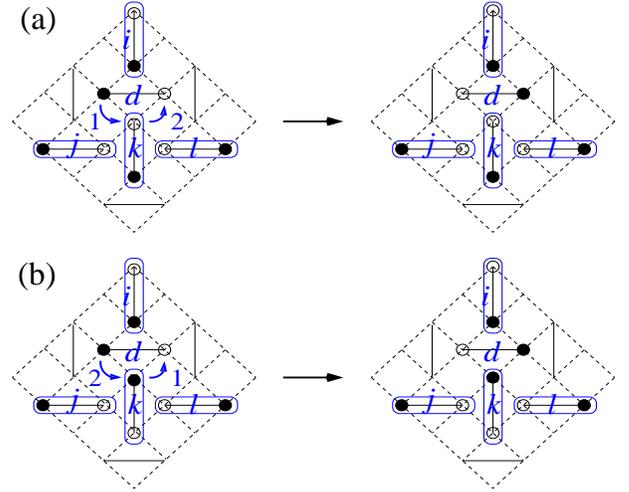}
\caption{(Color online)
Two different spin exchange processes contributing to the effective Hamiltonian $H^{(2)}_{eff}$ in second order perturbation theory.
These processes mediate the formation of an $S^z_{tot} = 0$ triplet state on the corresponding dimer,
while blocking  its formation on the neighboring dimers (inside ellipses).
Full (open) circles denote spin up (down) states. Numbers indicate the order of the spin exchange in the processes.
}\label{fig:secondorderperturbation}
\end{figure}

In order to assess the nature of these correlations, we consider processes occurring in second order perturbation theory.
These involve two spin exchanges along axial bonds. Apart from diagonal terms,  the result of such processes is again to flip
the effective spin on one of the  dimers, as shown in Fig.~\ref{fig:secondorderperturbation}.
The matrix element of each such  process depends  in detail on the specific local spin configuration on the dimers neighboring the dimer that undergoes the spin flip. Among the various possibilities, one with the largest amplitude is shown in Fig.~\ref{fig:secondorderperturbation} (a).
In the second order effective Hamiltonian  $H^{(2)}_{eff}$, this process contributes a term
\begin{equation}
\label{eq:secondorderperturbation}
\frac{2(\Delta J)^2}{2J-J'}\widetilde{S}^x_{d}(\frac{1}{2}-\widetilde{S}^z_{i})(\frac{1}{2}-\widetilde{S}^z_{j})(\frac{1}{2}-\widetilde{S}^z_{k})(\frac{1}{2}+\widetilde{S}^z_{l}),
\end{equation}
which provides a further contribution to the transverse field operator at site $d$, dressed by diagonal operators from the neighboring dimers, that project out the specific configuration of  dimer spin states according to the configuration shown in Fig.~\ref{fig:secondorderperturbation}.
While the transverse field in $H^{(1)}_{eff}$ acts locally on each dimer, the dressed transverse field operators deriving from  $H^{(2)}_{eff}$ lead to correlations among nearest neighbor dimer spins. For example, the spin exchange process on dimer $d$ shown in
Fig.~\ref{fig:secondorderperturbation} (a) could not take place as indicated by the arrows, if dimer $k$ was in the opposite spin state. Instead,  a different process could take place, as shown in Fig.~\ref{fig:secondorderperturbation} (b), that leads to a similar term in $H^{(2)}_{eff}$, but with a different energy denominator than in Eq.~(\ref{eq:secondorderperturbation}). In this way, details of the local dimer configuration enter the effective Hamiltonian in a rather complex manner.

The explicit form of the total effective Hamiltonian up to second order in $\Delta$ involves several  terms,  containing products of up to five effective spin operators, such as the term given explicitly in Eq.~(\ref{eq:secondorderperturbation}).
While the ground state of this effective Hamiltonian is not directly accessible, the general structure of these terms indicate that it will be a dressed version of $|\psi_D\rangle$, with  local inter-dimer correlations induced by the above virtual spin exchange processes.
This  reflects the QMC result, that the true ground state is close in energy to  $|\psi_D\rangle$, and does not exhibit long ranged correlations.

\begin{figure}[t]
\centering
\includegraphics[width = 0.45\textwidth]{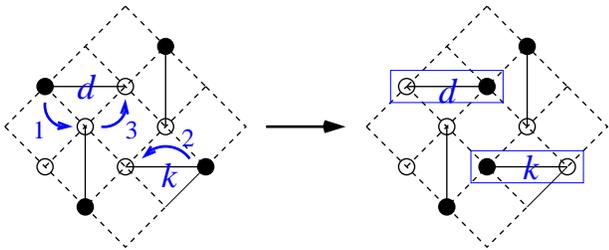}
\caption{(Color online) Spin exchange process contributing to the effective Hamiltonian $H^{(3)}_{eff}$ in third order perturbation theory. This process
leads to a flip of the effective spins on two neighboring dimers (inside boxes). As a result of this process, the effective spins inside the boxes have been exchanged. Full (open) circles denote spin up (down) states. Numbers indicate the order of the spin exchange processes.}\label{fig:thirdorderperturbation}
\end{figure}

The effective Hamiltonian up to second order in perturbation theory does not include transverse spin exchange terms, that would contain products of off-diagonal effective spin operators such as
$\widetilde{S}^+_{d}\widetilde{S}^-_{k} + \widetilde{S}^-_{d}\widetilde{S}^+_{k}$ on two neighboring dimers $d$ and $k$. However, such terms  appear in third order of the degenerate perturbation theory, e.g. from the process shown in
Fig.~\ref{fig:thirdorderperturbation}.
Similar to the transverse magnetic field operator in second order,
the matrix elements depend on the details of the spin configuration on the neighbors of the two dimers that undergo an effective spin exchange. The transverse effective spin exchange operators are thus similarly dressed by additional projection operators. Becoming more relevant at large values of $\Delta$, we expect that such exchange terms eventually drive the transition from the dimer triplet phase to the superfluid region, that was observed in the QMC simulations (c.f. Fig.~\ref{fig:groundstatephasediagram}).

We now turn to the special point $J'=2J$, where the ground state in the Ising limit has an enhanced degeneracy, as  discussed in Sec. II. Consider one of the $J'$-dimer of the Shastry-Sutherland lattice in the Ising limit $\Delta=0$. For $J'=2J$, this dimer can be either in one of the local configurations of Fig. 2 (a), which are also allowed configurations for $J'>2J$, or it is part of one of the local configurations shown in  Fig. 2 (b). Introducing a finite $\Delta>0$ in first order perturbation theory,
a splitting in the  energy between the configurations of Fig. 2 (a) and those of Fig. 2 (b) results, since the configurations of Fig. 2 (b) cannot gain exchange energy by a transverse spin exchange along the dominant $J'$ bonds, in contrast to the configurations of Fig. 2 (a).
To first order in $\Delta$,  this energy difference equals
\begin{equation}
 \delta E = -J +\frac{J'}{2}+\frac{\Delta J'}{2},
\end{equation}
near the point $(\Delta, J'/J)=(0,2)$.
A finite $\delta E>0$ thus leads to a partial lifting of the ground state degeneracy, and only the
local configurations of Fig. 2 (a) remain to span the low-energy sector. The configurations of Fig. 2 (b) are split-off by a finite energy difference $\delta E>0$ from the ground state manifold.
Since this energy difference remains positive for
\begin{equation}
 \delta E>0 \Longleftrightarrow \frac{J'}{J}>\frac{2}{1+\Delta}  
\end{equation}
we expect that as long as $J'/J\gtrsim 2/(1+\Delta)$ close to $(\Delta, J'/J)=(0,2)$, the system is driven by quantum effects towards the same phase as for $J'>2J$. In fact, this expectation is in agreement with the
QMC phase diagram  of Fig. 3, where  we found that (i) for small $\Delta>0$ at $J'=2J$, the system enters the dimer triplet state as it does for $J'>2J$, and (ii) the phase boundary of the dimer triplet phase closely follows the limiting line $J'/J= 2/(1+\Delta)$
according to $\delta E=0$ for small $\Delta$ (compare to the dashed-dotted line in Fig. 3).
The above argument was based on energy considerations on an isolated dimer.  Due to this restriction, we are not able here to
discern, if
the superfluid phase indeed terminates at a small, but finite values of $\Delta$, or if it persists down to any finite value of $\Delta>0$.
This would require the intra-dimer exchanges to be taken into account within higher orders of  perturbation theory. However, from our analysis of the case $J'>2J$ discussed above, we expect that  also in this case, the effective model will not allow for an explicit solution, thus leaving this question unanswered.
Hence, here we do not attempt to extend on this issue, but instead move on to study the magnetization process in the model under consideration.

\section{Magnetization Process}
After exploring  the ground state phase diagram, we now consider the model of Eq.~(1) in the presence of a finite magnetic field $h$, which couples to the spins by the standard Zeeman term,
\begin{equation}
 H\rightarrow H-h\sum_i S^z_i.
\end{equation}
From considering a singly flipped  spin with respect to the fully polarized state, one finds that the system is fully polarized for magnetic fields $h$ beyond
\begin{equation}
h_s=(2J+J'/2)(1+\Delta).
\end{equation}

Before discussing the magnetization process of the full quantum model, it is  again useful to consider first the
Ising limit.
The authors of Ref.~\onlinecite{siemensmeyer07} state that the Ising model on the Shastry-Sutherland lattice exhibits a magnetization process with a single plateau at $m/m_s=1/2$, where $m$ denotes the magnetization, and $m_s$ its saturation value, at least for $J'<2J$. In our notation,
this $1/2$ plateau should extend between  $2 J - J'/2 < h < 2J +J'/2$. In particular, in case $J'=J$, well within the N\'eel ordered zero-field regime, this range becomes $3/2< h/J < 5/2$. This conclusion in Ref.~\onlinecite{siemensmeyer07} was based on analyzing a finite system with 16 spins only. In order to check this conclusion on larger system sizes, we
performed a  systematic finite size analysis, using classical Monte Carlo (MC) simulations on lattices with up to
$18\times18$ spins in the canonical ensemble. For these simulations, we employed a single spin-flip Metropolis algorithm, and allowed for a simulated annealing of the system from a large initial temperature $T\approx J$ down to the final temperature during an initial state of equilibration. This way we were able to obtain MC results down to $T/J=0.1$ for $L=4,6$, and  $T/J=0.4$ for $L=18$. While for the purpose of the current study,  one can draw relevant conclusions on the magnetization process from these data, it will be interesting to obtain more refined numerical data on the magnetization process in the Ising limit, using extended ensemble sampling methods, similar to the approach taken in Ref.~\onlinecite{hwang07} for the square and triangular lattice. However, this lies beyond the scope of the current study, which is directed towards the quantum regime.
\begin{figure}[t]
\centering
\includegraphics[width = 0.45\textwidth]{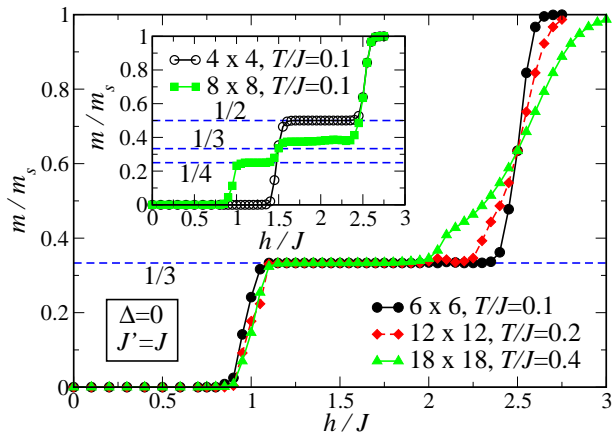}
\caption{(Color online) Magnetization curves $m/m_s$ as functions of $h$ for  the Ising model on the Shastry-Sutherland lattice at $J'=J$ for
different system
sizes, obtained from Monte Carlo simulations. The inset shows the results for the $4\times 4$ and $8\times 8$ lattices. }\label{fig:magisinga}
\end{figure}

Upon presenting the MC results,
we first discuss the  case $J'=J$ considered in Ref.~\onlinecite{siemensmeyer07}. Our MC data for the magnetization process on  different lattices are collected in Fig.~\ref{fig:magisinga}. The inset of
Fig.~\ref{fig:magisinga}  shows our  data for a $4\times4$ system, which appear to confirm the conclusion of Ref.~\onlinecite{siemensmeyer07}. However,  upon increasing the system size, we find that different plateau structures appear. In general,  in order to allow a system to establish a certain magnetization plateau, appropriate lattice sizes and boundary conditions must to be chosen. Otherwise, geometric constraints could frustrate certain magnetization patterns. In the present case, it can be seen from the numerical data, that only for  $L$ a multiple of 3, the system establishes a wide $m/m_s=1/3$ plateau, which is consistently observed for $L=6, 12$ and $18$. Instead, the $L=4$ and $8$ systems cannot establish the corresponding magnetic superstructure, and hence
lead to rather different magnetization curves with strong finite size effects, that do not represent  thermodynamic limit behavior.
The magnetization curve in Fig.~\ref{fig:magisinga} for $L=6$, taken at $T/J=0.1$, exhibits a magnetization plateau at $m/m_s=1/3$, extending from $h=J$ up to the saturation field at $h=5/2 J$. The data shown for the two larger systems, for which higher temperatures had to be taken in the MC simulations, are consistent with a thermal smothering of the magnetization jumps out of the plateau towards $m=0$ and full saturation, respectively.
We  conclude that for $J'=J$ the Ising model exhibits a single intermediate magnetization plateau at $m/m_s=1/3$, extending from
$h=J$ up to the saturation field at $h=5/2 J$. The different conclusion of Ref.~\onlinecite{siemensmeyer07} appears to be due to the usage of inappropriate finite lattice sizes. Note, that all values of  $L$ considered here were even, i.e. a plateau at $m/m_s=1/2$, if it would exist in this model, would not be frustrated by finite lattice effects. In fact, as discussed below, we find that such a $1/2$ plateau appears for finite values of $\Delta$ due to quantum effects.
\begin{figure}[t]
\centering
\includegraphics[width = 0.45\textwidth]{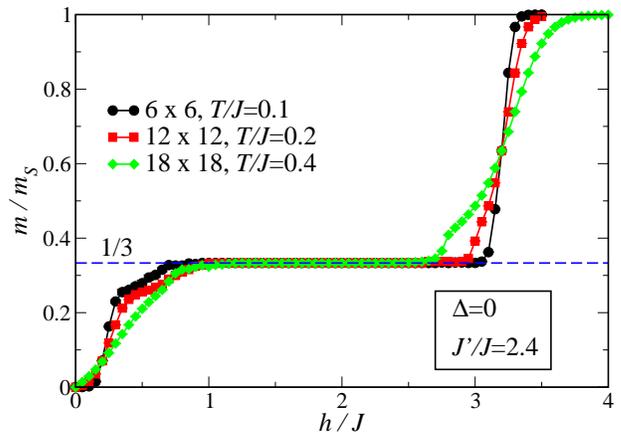}
\caption{(Color online) Magnetization curves $m/m_s$ as functions of $h$ for the Ising model on the Shastry-Sutherland lattice at $J'=2.4J$ for
different system sizes, obtained from Monte Carlo simulations.}\label{fig:magisingb}
\end{figure}
Next, we consider the case $J'=2.4J$, well inside the degenerate region of $J'>2J$. Below, we will compare the Ising model result to QMC data on the magnetization process for finite $\Delta>0$ at the same value of $J'/J$. The MC results for the magnetization process in the Ising limit are shown in Fig.~\ref{fig:magisingb},
where we  now consider linear system sizes $L$ that are a multiple of $6$. Again, we observe a wide $1/3$ magnetization plateau, which extends from $h\approx 0.8J$ up to the saturation field at $h=h_s=3.2 J$. A precise estimate of the lower boundary of the plateau is not accessible from the current finite size data, as we could not collect data at sufficiently low temperatures on  larger  systems. There appears however a finite magnetization with a smooth increase well before
the plateau is entered. It will be interesting to explore this low$-m$ regime in more detail using extended ensemble methods. The point that is important for the following discussion, and which follows also from the current MC data, is the absence again of a magnetization plateau at $m/m_s=1/2$. Instead, the magnetization exhibits a jump from the  $1/3$ plateau up to magnetic  saturation at $h=h_s$.

\begin{figure}[t]
\centering
\includegraphics[width = 0.45\textwidth]{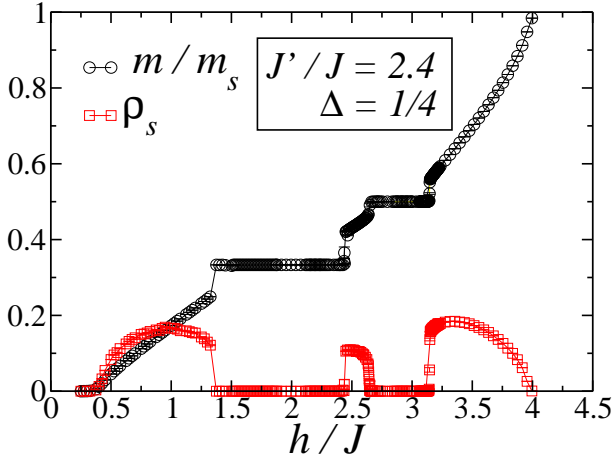}
\caption{(Color online)
Magnetization $m/m_s$ and spin stiffness $\rho_S$ for $J'=2.4J$ and $\Delta=1/4$ as functions of the applied magnetic field $h$.
}\label{fig:magquantum}
\end{figure}
\begin{figure}[t]
\centering
\includegraphics[width = 0.45\textwidth]{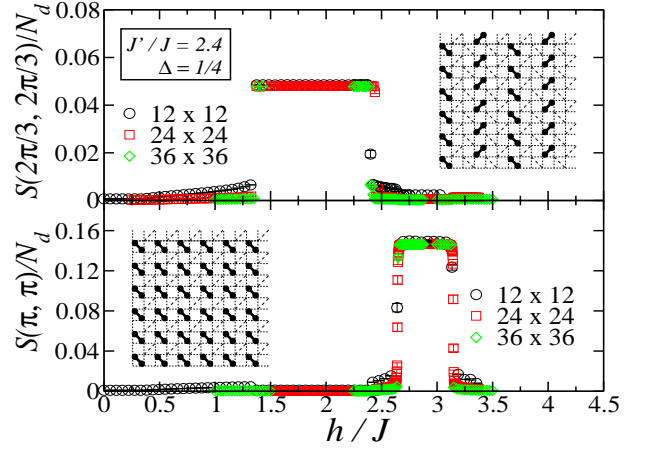}
\caption{(Color online)
Dimer super structure factors $S(2\pi/3,2\pi/3)$ (upper panel) and $S(\pi,\pi)$ (lower panel) at $J'=2.4J$ and $\Delta=1/4$ as functions of the applied magnetic field $h$. The insets illustrate the corresponding magnetic superstructures, where a dumbbell denotes a fully polarized dimer. $N_d=N/2$ is the number of dimers.
}\label{fig:magsuperstructures}
\end{figure}

As we  show next,
a plateau  at $m/m_s=1/2$,  while not obtained in the Ising limit, emerges in the quantum model at finite values of $\Delta>0$.
Fig.~\ref{fig:magquantum} shows the magnetization process for  $J'=2.4J$ and $\Delta=1/4$ from  QMC simulations. We again find a magnetization plateau at $m/m_s=1/3$ as well as a  plateau at $m/m_s=1/2$. Adding a transverse exchange to the Ising model thus leads to a
softening of the large magnetization jump from $m/m_s=1/3$ to $1$ in the Ising limit, and an additional plateau region appears with $m/m_s=1/2$.
In hard-core bosonic language, the
transverse exchange maps onto a non-frustrating hopping amplitude. In the current situation, this finite boson hopping does not lead always to
superfluidity, but drives the system into an insulating phase at filling $\rho=3/4$, corresponding to $m/m_s=1/2$ (due to  particle-hole symmetry, a similar insulating region emerges also for  $\rho=1/4$, corresponding to $m/m_s=-1/2$).
Indeed, we find from Fig.~\ref{fig:magquantum}, that the spin stiffness $\rho_s$ vanishes inside both plateau regions. It is then in order to study, if the magnetic excitations in both plateau phases form
period crystals, and what the structures of such solid arrays would be. Long-range crystalline ordering of magnetic excitations inside  magnetic plateau regions has previously been analyzed for the isotropic spin-1/2 Heisenberg model on the Shastry-Sutherland model~\cite{momoi00a,momoi00b,miyahara00,mishguich01,miyahara03b,sebastian07,dorier08,abendschein08}, based on different approximative schemes.
The relevant structures expected from these studies are shown in the inset of Fig.~\ref{fig:magsuperstructures} for the $1/3$ (upper panel), and the $1/2$ (lower panel) plateau, respectively.
They are  expressed in terms of periodic arrangements of  $S^z_{tot}=1$ dimer triplet states $|\uparrow\uparrow\rangle_d$, denoted by  dumbbells, in  a background of less polarized, e.g.  $S^z_{tot}=0$ states on the remaining dimers. We now assess, if these crystalline patterns  appear also in  the current quantum model, which represents an anisotropic version of the Heisenberg model considered thus far.

\begin{figure}[t]
\centering
\includegraphics[width = 0.45\textwidth]{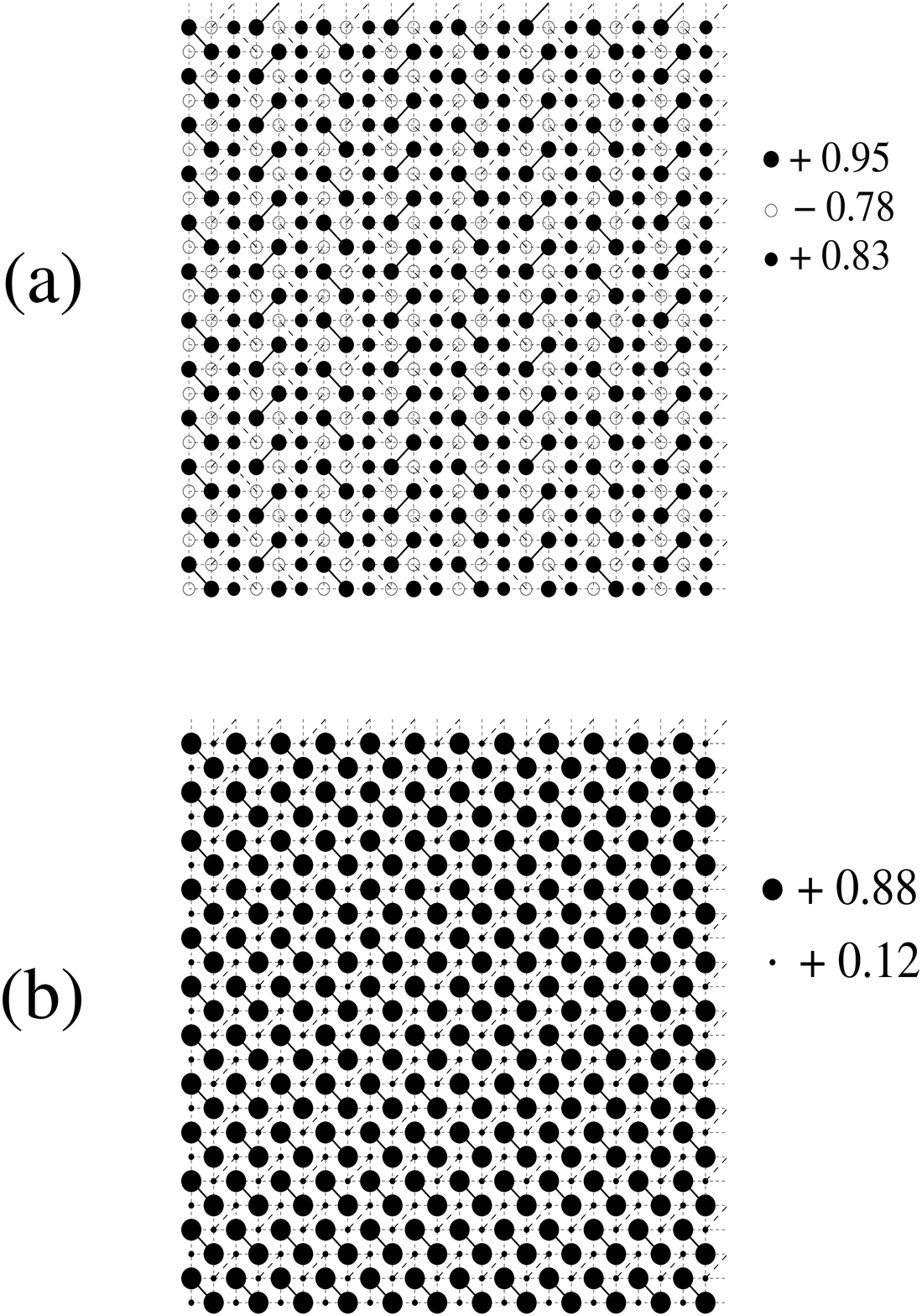}
\caption{(Color online)
Distribution of the local magnetization $m_i/m_s$ for the magnetization plateaus at $m/m_s=1/3$ (a) and 1/2 (b), represented as indicated by the size of circles on the lattice sites. Results are shown for a $24\times 24$ lattice.
}\label{fig:realspace}
\end{figure}

In a QMC simulation,  one can probe exactly for these magnetic superstructures  by measuring an appropriate structure factor. Hence, we analyze the ordering pattern of the magnetic excitations by measuring the triplet excitation structure factor
\begin{equation}
S(\mathbf{q})=\frac{1}{N_d} \sum_{d,k} e^{i\mathbf{q}(\mathbf{r}_d-\mathbf{r}_k)} \langle P_d P_k \rangle,
\end{equation}
where $P_d$ is a projector on the $S^z_{tot}=1$ dimer triplet state $|\uparrow\uparrow\rangle_d$ on dimer $d$. $N_d=N/2$ equals the number of dimers in the finite system of $N$ sites. The positions $\mathbf{r}_d$ of the dimers and the momentum space vector $\mathbf{q}$ are defined with respect to the convenient coordinate system formed by the square lattice of dimers, with the distance between the centers of two neighboring dimers taken as unity. In this way, the solid order shown in the upper panel of Fig.~\ref{fig:magsuperstructures} for  $m/m_s=1/3$ corresponds to a peak in $S(\mathbf{q})\propto N_d$ at $\mathbf{q}=(2\pi/3,2\pi/3)$, and at $\mathbf{q}=(\pi,\pi)$ for the order shown in the lower panel of Fig.~\ref{fig:magsuperstructures} for  $m/m_s=1/2$.
The main panels of Fig.~\ref{fig:magsuperstructures} show these components of the structure factor for different system sizes, respectively. We checked, that no  other signal in $S(\mathbf{q})$ appeared, apart from those explicitly shown.
From these data we conclude, that indeed the crystalline orders of the magnetic excitations shown in  Fig.~\ref{fig:magsuperstructures} are stabilized within the two magnetization plateau regions, respectively. In order to study the  real-space distribution of the magnetization among the  lattice sites in more detail, we also measured the mean local values of the magnetization. The obtained distributions of the local magnetization are shown for the 1/3 and 1/2 plateau in Fig.~\ref{fig:realspace} (a) and (b), respectively. They agree well with the distributions reported for the isotropic Heisenberg model in the studies mentioned
above. 

\begin{figure}[t]
\centering
\includegraphics[width = 0.45\textwidth]{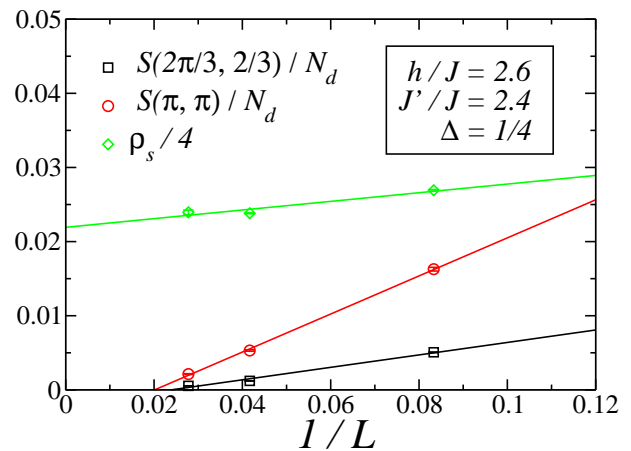}
\caption{(Color online)
Finite size scaling of the superfluid density $\rho_S$ and the dimer superstructure factors of the neighboring plateaus at $h=2.6$ for $J'=2.4J$ and $\Delta=1/4$.
}\label{fig:supersolid}
\end{figure}

Upon varying $h$ beyond these wide magnetization plateaus, the magnetization undergoes  discontinuous jumps, as seen in
Fig.~\ref{fig:magquantum}, alert when entering the $1/2$ plateau from below, where such a jump appears not that well resolved. The spin stiffness $\rho_s$ exhibits a similar behavior, with clear jumps upon entering the plateaus, expect when entering the $1/2$ plateau from below. Apart from this case, the superfluid-insulator transitions are thus clearly first order. Concerning the region below the $1/2$ plateau, one might instead consider the presence of a supersolid state of  magnetic excitations, with both a finite superfluid density and a crystalline superstructure. In order to assess, if supersolid behavior is indeed present in this regime, we show in  Fig.~\ref{fig:supersolid} the finite size scaling of both the superfluid density $\rho_s$ and the relevant structure factors from the neighboring magnetization plateaus at a magnetic field of $h=2.6$, inside the possibly supersolid region. From the finite size scaling we can exclude the presence of a supersolid phase in this parameter regime. Between the two magnetization plateaus, the system is thus in a uniform superfluid state.
Our analysis of the magnetization process at $J'/J=2.4$ and $\Delta=1/4$  did not exhibit the presence of any additional plateaus at lower values of $m$. Indeed, at  low-$m$ the magnetization curve as well as the superfluid density appear smooth in Fig.~\ref{fig:magquantum}. In particular, we did not find any of the fractional magnetization plateaus mentioned in Sec. I for the compounds SrCuB${}_2$(BO${}_3$)${ }_2$ and TmB${}_4$, nor those found for the isotropic spin-1/2 Heisenberg model on the Shastry-Sutherland model in recent studies~\cite{sebastian07,dorier08,abendschein08}. We postpone a discuss on this issue to the concluding section.

\begin{figure}[t]
\centering
\includegraphics[width = 0.45\textwidth]{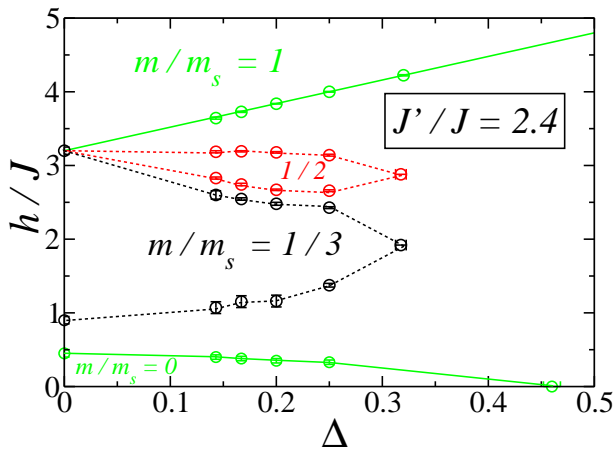}
\caption{(Color online)
Magnetic phase diagram of the  spin-1/2 XXZ model on the Shastry-Sutherland lattice with ferromagnetic transverse spin exchange at  $J'/J=2.4$ depending on the anisotropy  $\Delta$ and the applied  magnetic field $h$.
}\label{fig:emergence}
\end{figure}

Finally, we present our results for the magnetization process upon varying over a wider range of parameters space. The results of QMC simulations similar to those discussed in detail above are summarized in Fig.~\ref{fig:emergence}, which shows the magnetic phase diagram for a generic fixed ratio of $J'/J=2.4$. The figure shows both the $m/m_s=1/3$ magnetization lobe that extends from the Ising limit up to $\Delta=0.32(2)$, as well as the emergent $m/m_s=1/2$ lobe, that extends up to
a similar value of $\Delta$, but shrinks upon approaching the Ising limit, where no $1/2$ plateau persists.

\section{Conclusions}
We studied the ground state  phase diagram and the magnetization process of the spin-1/2 easy-axis XXZ model of  Eq.~(1) with ferromagnetic transverse spin exchange on the Shastry-Sutherland lattice. The model exhibits a N\'eel ordered phase for small values of $J'/J$ and $\Delta$, and a superfluid phase for dominant $\Delta$.
For sufficiently large values of $J'/J$, a dimerized phase with the dominant formation of $S_{tot}^z=0$ triplet states on the $J'$-dimer bonds is stabilized, that connects to the degenerate phase of the model in the Ising limit. We like to compare this findings to the case of the isotropic Heisenberg model on the same lattice.
Also in this model there is strong evidence for an intermediate phase~\cite{miyahara03a}, separating the  N\'eel ordered phase from the dimer singlet phase.
In contrast, this intermediate phase is  however not a superfluid, but appears to be realize a valence bond crystal  driven by  the frustrated nature of the transverse exchange interactions in this region~\cite{laeuchli02}.

Using classical Monte Carlo simulations to assess the magnetization process in the Ising limit, we found in contrast to previous claims~\cite{siemensmeyer07} no magnetization plateau at $1/2$ of  full saturation, but instead a $1/3$ plateau, with a jump towards full saturation. The  quantum model shows the presence of a $1/2$ plateau in addition to the $1/3$ plateau. This emergence of a magnetization plateau upon adding quantum fluctuations (via a finite $\Delta$) to the Ising model remains of the
previously observed emergence of a magnetization plateau via thermal fluctuations in the classical Heisenberg model on the kagome lattice~\cite{zhitomirsky02}. While we examined the case of a ferromagnetic transverse spin exchange,
the experimental observation of a $1/2$ plateau
in the compounds TmB${}_4$ and  SrCuB${}_2$(BO${}_3$)${ }_2$ can be taken as indication, that such a plateau  also emerges for antiferromagnetic transverse exchange. Indeed, a $1/2$ plateau is consistently reported also in the studies on the magnetization process of the isotropic Heisenberg model on the Shastry-Sutherland lattice~\cite{miyahara03a}.

With respect to other reported magnetization plateaus both from experiments and in recent theoretical work, we did not obtain
evidence from our quantum Monte Carlo simulations, that they are stabilized in the current model.
In particular, the pronounced $1/6$ plateau, consistently
reported in recent theoretical work~\cite{sebastian07,dorier08,abendschein08},
has  not been found for the parameter region we examined in our simulations (we performed various scans inside the range
$2<J'/J<3$, down to $\Delta=1/7$, with results similar to those presented in detail above),
even though the proposed magnetic unit cell~\cite{sebastian07,dorier08,abendschein08} is commensurate with the finite lattices that we employed in our simulations. We explicitly checked that no signal in the relevant component of $S(\mathbf{q})$ at $\mathbf{q} = (\pi, \pi/3)$ appears near $m/m_{s} = 1/6$ in this model. Hence, we conclude that the model
considered here exhibits magnetization plateaus at $1/2$ and $1/3$ of the full saturation, only.
Future work could explore in more detail the magnetization process in the Ising  model inside the low-field region
using extended ensemble methods, similar to the approach taken in Ref.~\onlinecite{hwang07}.
It would be interesting to apply the theoretical approaches employed in
Refs.~\onlinecite{dorier08} and \onlinecite{abendschein08} to the current model, in order to
to check their predictions against the unbiased large-scale
quantum Monte Carlo results present here.

\section*{Acknowledgments}
We thank A. L\"auchli, F. Mila, R. M\"ossner, A. Muramatsu and D. Poilblanc for useful discussions.
Financial support by the Deutsche Forschungsgemeinschaft
under Grant No. WE 3649/2-1 and through  SFB/TR 21 is gratefully acknowledged, as well as
the allocation of CPU time by HLRS Stuttgart and NIC J\"ulich.
The employed numerical simulation code was partially based on the ALPS libraries~\cite{albuquerque07b}.


\end{document}